# An Optimal Fully Distributed Algorithm to Minimize the Resource Consumption of Cloud Applications

Nikos Tziritas[1], Cheng-Zhong Xu[1,3], Jue Hong[1]
[1]Chinese Academy of Sciences, China
{nikolaos, cz.xu, jue.hong}@siat.ac.cn

Samee Ullah Khan[2]
[2]North Dakota State University, Fargo, ND, USA
samee.khan@ndsu.edu

[3]Wayne State University, Detroit, MI, USA
czxu@wayne.edu

*Abstract—* **According to the pay-per-use model adopted in clouds, the more the resources consumed by an application running in a cloud computing environment, the greater the amount of money the owner of the corresponding application will be charged. Therefore, applying intelligent solutions to minimize the resource consumption is of great importance. Because centralized solutions are deemed unsuitable for large-distributed systems or large-scale applications, we propose a fully distributed algorithm (called DRA) to overcome the scalability issues. Specifically, DRA migrates the inter-communicating components of an application, such as processes or virtual machines, close to each other to minimize the total resource consumption. The migration decisions are made in a dynamic way and based only on local information. We prove that DRA achieves convergence and results always in the optimal solution.**

## I. INTRODUCTION

In the last few years, there has been a great effort from the cloud providers to offer user friendly environments that can be used by clients to execute applications. The services are delivered to the cloud users through a pay-per-use-model, which means that the owner of an application is required to pay an amount of money that is (approximately) proportional to the amount of resources the respective application consumes during its execution on the cloud. Therefore, applying intelligent techniques to minimize the resource consumption is of paramount importance.

The aforementioned problem can be solved by identifying an assignment scheme between the interacting components, such as processes and virtual machines of an application, and the computing nodes of a cloud system, such that the total amount of resources consumed by the respective application is minimized. However, the problem is NP-complete for general network graphs, even in the case of unbounded capacity ([1], [8]). Therefore, many researchers have focused on some variations of the network graph, such as homogeneous arrays [5] or trees [6], which facilitates in solving the problem in polynomial time.

The above mentioned problem becomes even more interesting when considering applications that are prone to load changes. This is due to the fact that in such kinds of problems, an assignment scheme may be optimal for some time interval, while sub-optimal for some time interval. Therefore, it is meaningful to dynamically reassign the processes to nodes taking into account the new communication demands of the application components. However, extra care must be taken to not make redundant calculations for processes that are already optimally located within the system. Therefore, approaches that build on the min-cut max-flow technique, such as reported in [5] and [6] become inapplicable when considering large-scale applications, because of their extraneous demands of both memory and execution time.

In this work, we aim to solve the following problem: *Assuming that an application is already hosted in a cloud, reassign the application components to nodes of the system so as to minimize the total communication and computational resources consumed by the components of the respective applications.*

We approach this problem by proposing a fully distributed algorithm that leverages on the min-cut max-flow technique. It should be stressed that the min-cut algorithm is invoked only for very small portions of the application graph. The main idea behind our algorithm is that each node must identify, based only on local information, which processes or group of processes (hosted by the respective node) must migrate to other locations to minimize the resources consumed by an application. Because the studied problem is intractable for networks that are modeled as general graphs, we restrict the optimality proof of our algorithm to tree-structured networks. However, we extend the functionality of our proposed algorithm to run also on hierarchical networks. In this paper, we also provide a proof about DRA's convergence. The innovation of our work lies in the fact that this is the first time a problem of such nature has been solved in a distributed fashion, leading at the same time to the optimal solution.

This paper is structured as follows: Section II reports the related work; in Section III we describe the system model and rigorously formulate the optimization problem; the proposed algorithm is detailed in Section IV; in Section V we provide a thorough discussion about the convergence and optimality proof of our algorithm; while Section VI concludes the work.

## II. RELATED WORK

As discussed in the preceding section, numerous works have been proposed to tackle the aforementioned problem in a centralized way. The most recent ones are discussed in the following text. Initially the problem was tackled for a dual processor system [2], [15]. A great deal of works focused on extending the above solutions to address the case of a fully-connected network of *n*-processors [1], [8], [9], [19]. Researchers focused also on various structures of graphs, such as tree, linear array, and bipartite graphs to solve the problem optimally [5], [6], [18]. Recently, the above problem has also been tackled as a scheduling problem, with the target function being the minimization of both the makespan and either energy consumption [20] or communication overhead [4], [11]. Other recent works target bi-criteria problems for the MultiProcessor System-on-Chip (MPSoC) architectures [13], [21], [22], [23]. The problem tackled in this paper is also of interest in the field of data engineering; with the goal being to minimize the total communication overhead while executing a large number of queries simultaneously in a distributed environment [7], [12].

Over the last few years, large-scale distributed systems (e.g., clouds, grids, sensor networks etc) gained a lot of attention. Centralized solutions were rendered useless to these systems due to scalability issues, which finally were superseded by distributed solutions. In the following paragraphs we discuss the most recent distributed algorithms related to this problem. In [10], the authors propose a decentralized algorithm to minimize the communication overhead produced by query operators located in large-distributed systems. Their algorithm takes advantage of the spring relaxation technique to place the query operators in such a way so as to minimize the total network traffic produced by them. The application and the network are modeled as general graphs. Ref. [16] and Ref. [17] propose fully distributed algorithms that migrate agents in wireless sensor networks to minimize the total communication cost. The algorithms migrate agents towards the center of gravity of their communication load. Both the application and the network of these works are restricted to tree-structured graphs. Another work that is closer to our topic is that of Sonnek *et al* [14] that considers virtual machine migrations for minimizing the total communication overhead. They propose a *distributed bartering algorithm* that allows virtual machines (VMs) to negotiate placement on a physical server that is closer to the data they require. They consider general application graph and hierarchical network topologies. The drawback of the aforementioned work is that their algorithm considers migrating the application components as single entities without taking into account groups of mutually dependent components (communicating heavily with each other).

Our work must be considered unique compared to all of the above referenced work in:
- Tackling the problem in a fully dynamic and distributed manner.
- Identifying the inter-dependencies between the components of the application graph.
- Proving that our algorithm results always in the optimal solution.
- Guaranteeing that the proposed methodology always converges.
- Making only local decisions, incurring in that way negligible overhead in terms of the system control messages.
- Developing a methodology that can equally operate well trees and hierarchical networks.
- Handling the cases where the nodes within the system are overloaded.

## III. SYSTEM MODEL AND PROBLEM FORMULATION

### A. System Model

Initially, we assume that our network comprises of $N$ computing nodes structured as a tree, while in the sequel we expand the scope to capture hierarchical topologies (with each node representing a cluster of connected servers). Let $n_x$ be the $x^{th}$ node in the network, while $h_{xy}$ captures the path length between $n_x$ and $n_y$ ($h_{xx} = 0$). Let $P$ be the number of the processes comprising an application to be executed, and $p_i$ define the $i^{th}$ such process. The communication dependencies between processes are encoded by an $P \times P$ matrix denoted by $C$, with $c_{ik}$ representing the amount of data (in bytes) sent by $p_i$ towards $p_k$, while "$c_{ik} + c_{ki}$" reflecting the amount of data exchanged between $p_i$ and $p_k$ ($c_{ik} \neq c_{ki}$). Let the execution cost of $p_i$ be known *a priori* and captured by the expression $u_i > 0$. The above are fully compatible with the models adopted in [5], [6], [14], and Amazon EC2 [24].

To capture the communication dependencies between processes (or Virtual Machines (VMs)) and nodes, we extend the number of processes by $N$ virtual ones, with each of them representing a node. Therefore, the data exchanged between a virtual process and a real process represents the communicational demands between the real process and the node represented by the virtual process. Specifically, $p_i$ is a real process iff $0 < i \leq P$, and a virtual one iff $P < i \leq P + N$. By doing so, we are able to extend the matrix encoding the communicational dependencies between processes to an $(P+N) \times (P+N)$ matrix that includes the communicational dependencies between processes and nodes (besides the ones between processes). For simplicity, let the virtual process representing $n_x$ be denoted by $p_{n_x}$.

Figure 1 depicts an example of an application graph, capturing the communication and computational demands of the processes. Specifically, a white rectangle represents a real process, while a black one a virtual process. An edge reflects the communication dependencies between its incident processes, while the number pictured besides an edge signifies the amount of data exchanged between these processes (called weight of the respective edge). Note that in our example, $p_1$ communicates with two virtual processes ($p_{n1}$ and $p_{n2}$), with that meaning that $p_1$ has communication dependencies with both $n_1$ and $n_2$. Moreover, the number depicted next to a real process represents the cost of executing the respective process on a node.

## B. Problem Formulation

The problem is formally stated as: *given a network of N nodes that host an application of P processes, reassign the processes to nodes such that the application resource utilization is minimized.*

As discussed earlier, the application resource usage depends on both the execution and communication costs incurred by the processes belonging to the respective applications. According to the previous notations, the execution cost of an application can be directly calculated by summing the computational demands of each process. However, we cannot directly add the data exchanged between the processes to calculate the communication cost. This is because the communication resource usage is directly connected to the links involved for the communication between processes. Therefore, two cases arise for the communication between any pair of processes: **(a)** The processes are located on the same node. Because the inter-process communication is performed by accessing the local memory, the resource usage is negligible; and **(b)** The processes are located on different nodes. In this case, the resource usage is proportional to both the amount of data exchanged and the number of links involved for establishing the communications. The aforementioned model is used by many cloud providers, such as Amazon EC2 to charge the total communication resource usage of an application, which is proportional to the amount of data travelled over the network [25]. By way of example, the communication resource usage between $p_i$ and $p_k$ is equal to $D \times h_{xy}$, given that $p_i$ and $p_k$ exchange $D$ data and they are located on nodes $n_x$ and $n_y$, respectively. In the case of local communication we have $h_{xx}=0$, which entails that the communication cost is zero.

To express the problem through a rigorous mathematical formulation, we must define the following. Let $F$ be an $P \times N$ matrix capturing the assigned node for each process, with $f_{ix} = 1$ if $n_x$ is assigned to $p_i$, otherwise $f_{ix} = 0$. Therefore, given an assignment $F$, the total execution resource usage is given by *execRs(F)*, as described in Eq. 1, while the total communication resource usage (also called the *communication* or *network cost/overhead*) is denoted by *commRs(F)*, as described in Eq. 2. From the above, we can deduce that the total resource usage can be represented by the function *totalRs(F)*, which is expressed by Eq. 3. Therefore, the total resource usage can be minimized by finding an assignment $F^*$ such that Eq. 3 is minimized.

$$execRs(F) = \sum_{i=1}^{P} \sum_{x=1}^{N} u_i f_{ix} \quad \text{Eq. 1}$$

$$commRs(F) = \sum_{i=1}^{P+N} \sum_{k=1}^{P+N} \sum_{x=1}^{N} \sum_{y=1}^{N} c_{ik} f_{ix} f_{ky} h_{xy} \quad \text{Eq. 2}$$

$$totalRs(F) = execRs(F) + commRs(F) \quad \text{Eq. 3}$$

It can be seen that the execution cost represented by Eq. 1 does not depend on $F$ (constant). This means that the minimization of Eq. 2 must result in the optimal solution. Therefore, the execution cost is decoupled from our algorithmic design.

## IV. DISTRIBUTED REASSIGNMENT ALGORITHM (DRA)

By designing an algorithm that emphasizes on performing light-weight calculations, we do not put the solvability of our problem at risk. This is based on our central idea that we must avoid running into feasibility issues, such as trying to find the min-cut of a graph that does not even fit into the main memory for simulation purposes. Therefore, we focus on mechanisms that migrate a process or a small group of processes from one node to another, aiming at the total communication cost reduction. In the first section, we describe the single process migration mechanism. A significant drawback of the abovementioned mechanism is identified in the second section, leading that mechanism to make sub-optimal decisions. As a remedy to that drawback, we introduce the super-process migration mechanism. In the last section, we provide some implementation details.

### A. Single Process Migration Mechanism

Our objective is to perform migrations in such a way that further reduces the current network cost. In this section, we consider migrating processes as singular entities. Therefore, we need a metric to consider whether such a migration contributes negatively or positively to the minimization of the total network cost. To define such a metric, we first need to introduce some extra but necessary notations.

The variable $q_{ik}^z = 1$ iff $n_z$ is used for communication between $p_i$ and $p_k$, else $q_{ik}^z = 0$. In the special case where $p_i$ and $p_k$ are co-located, then we have $q_{ik}^z = 0$, irrespective of whether $n_z$ is used for their communication or not. Let $M_{sd}^i$ define a migration of process $p_i$ (called *target process*) from the hosting node $n_s$ (called *local node*) towards a 1-hop neighbor $n_d$ (called *destination node*). For any such migration, we need to identify the following:

*Positive load $pl_{sd}^i$*: This load represents the gain (in terms of the total communication overhead) when migrating $p_i$ from $n_s$ towards its 1-hop neighbor $n_d$. Specifically, this migration will bring $p_i$ nearer by 1 hop to the set of processes (hereafter referred as $A$) that use the destination node $n_d$ (as either a hosting or routing node) to communicate with $p_i$ when the latter is located on $n_s$. Therefore, the total communication overhead will decrease by an amount that is equal to the volume of data exchanged between $p_i$ and the processes belonging to $A$ (see Eq. 4).

*Negative load $nl_{sd}^i$*: This load represents the cost (in terms of the total communication overhead) when migrating a process $p_i$ from $n_s$ towards its 1-hop neighbor $n_d$. Specifically, when migrating process $p_i$ from $n_s$ towards $n_d$, $p_i$ will distance itself by 1 hop from the set of processes (hereafter referred as $B$) that do not use the destination node $n_d$ to communicate with $p_i$ when the latter is located on $n_s$. Therefore, the total communication overhead will increase by an amount that is equal to the volume of data exchanged between $p_i$ and the processes belonging to $B$ (see Eq. 5).

*Migration benefit $b_{sd}^i$*: This represents the metric that assesses whether a migration is beneficial or otherwise. Specifically, the migration of process $p_i$ from node $n_s$ to node $n_d$ is considered *beneficial* if the load $pl_{sd}^{i}$ is greater than the load $nl_{sd}^{i}$, otherwise it should be considered *non-beneficial*. The benefit $b_{sd}^{i}$ is expressed by Eq. 6 that states that the migration $M_{sd}^{i}$ will incur a decrease or an increase in the overall system communication overhead by an amount that is equal to the subtraction of $nl_{sd}^{i}$ from $pl_{sd}^{i}$. If the resultant of the subtraction is zero, then $M_{sd}^{i}$ will not affect the overall communication overhead. In case the result is a positive value, then the overall communication overhead will decrease by an amount equal to that value, otherwise it will be increased.

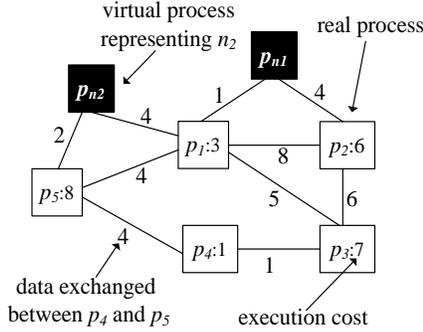

Figure 1. Application graph.

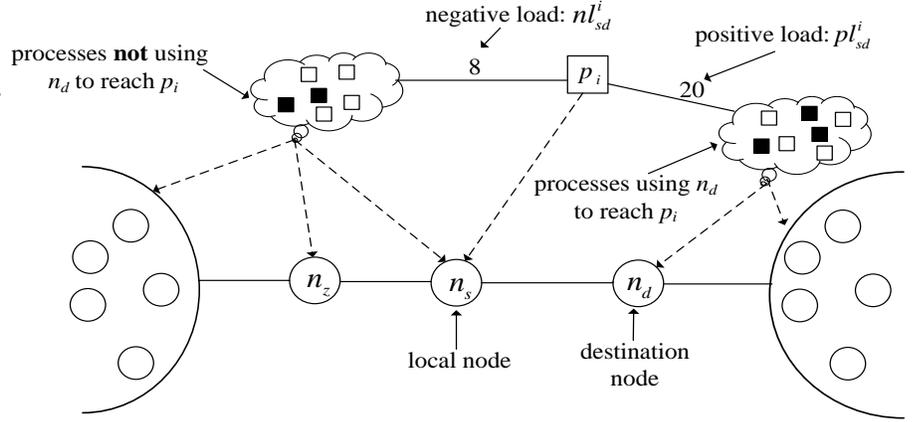

Figure 2. Tentative migration $M_{st}^{i}$ along with its positive and negative loads.

$$pl_{sd}^{i} = \sum_{k=1}^{P+N}(c_{ik}+c_{ki})q_{ik}^{d} \qquad \text{Eq. 4}$$

$$nl_{sd}^{i} = \sum_{k=1}^{P+N}(c_{ik}+c_{ki})(1-q_{ik}^{d}) \qquad \text{Eq. 5}$$

$$b_{sd}^{i} = pl_{sd}^{i} - nl_{sd}^{i} \qquad \text{Eq. 6}$$

*Center of gravity*: A process is said to be at the center of gravity of its communication load iff there is no possible migration of that process to any 1-hop neighbor (against the current hosting node) that leads to network cost reduction. Otherwise, the process is considered *unbalanced*.

We set out an example to illustrate the aforementioned definitions (see Figure 2). Consider a tentative migration of $p_i$ from $n_s$ to $n_d$, where according to the preceding text, $p_i$ plays the role of the target process; while $n_s$ and $n_d$ the role of local and destination node, respectively.

- Positive load $pl_{sd}^{i}$: The cloud at the right side of Figure 2 represents a group of processes that use the destination node $n_d$ as either a hosting or routing node to reach $p_i$. The amount of data exchanged between $p_i$ and the aforementioned processes represents the positive load $pl_{sd}^{i}$ that is equal to 20.

- Negative load $nl_{sd}^{i}$: The cloud at the left side of Figure 2 represents a group of processes that do not use the destination node $n_d$ to reach $p_i$. The amount of data exchanged between $p_i$ and the above mentioned processes represents the negative load $nl_{sd}^{i}$ that is equal to 8.

- Migration benefit $b_{sd}^{i}$: According to the definition of migration benefit and Eq. 6, the migration $M_{sd}^{i}$ will decrease the total communication cost by 12 = 20-8 = $b_{sd}^{i}$.

### B. Super-process Migration Mechanism

Performing only single process migrations may lead to sub-optimal solutions. Consider the example shown in Figure 3, where the processes $p_1$ and $p_2$ are hosted by $n_1$. As we can see, the current network cost is equal to 6, the benefit of migrating $p_2$ towards $n_1$ is equal to $-2$, while the benefit of migrating $p_1$ to $n_2$ is equal to $-9$. Moreover, it also can be observed that each process is at the center of gravity of its communication load. However, if both processes $p_1$ and $p_2$ migrate towards $n_2$, then the total network cost will be equal to 1, which is the optimal solution. Therefore, we need to find a methodology to identify groups of processes that their migrations lead into further network cost reduction. Before proceeding with the identification of such groups of processes, we first need to redefine Eq. 4, Eq. 5, and Eq. 6 to express a group of processes instead of a single process. To keep the analysis simple, below we detail a technique that transforms a group of processes into a super-process.

*Super-process:* According to the following procedure, a group of co-located *real* processes can be transformed into a super-process. The transformation starts by removing all of the edges that connect processes belonging to the targeted group. The above edges are also called *internal edges*, while the edges that connect processes of the targeted group with processes not belonging to that group are called *external edges*. All the external edges of the targeted group become edges of the super-process to be formed. In case the super-process, in question, acquires more than one edge towards an external process, then we merge all of these edges into one, with its weight being equal to the sum of the weights of the merged edges (see Figure 4 for further details).

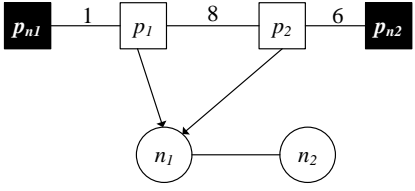
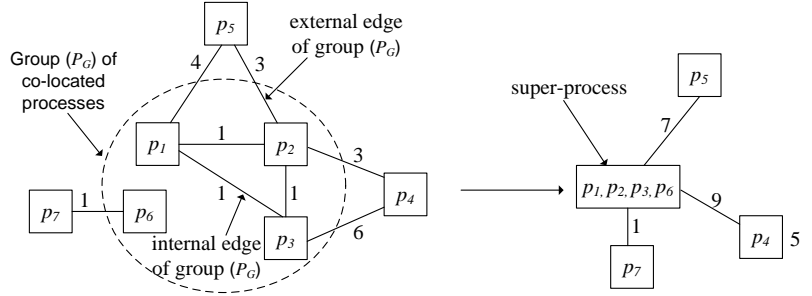

Figure 3. A sub-optimal case when considering only single process migrations.

Figure 4. Transforming a group of co-located processes into a super-process.

In the sequel, we give the mathematical equations capturing the positive load, negative load, and migration benefit for a tentative migration $M_{sd}^G$ of a super process $P_G$.

*Positive load $PL_{sd}^G$*: This represents the volume of data exchanged between $P_G$ and the processes using $n_d$ as either a hosting or routing node to reach one of the processes represented by $P_G$ (expressed by Eq. 7).

*Negative load $NL_{sd}^G$*: As expressed by Eq. 8, represents the volume of data exchanged between a process belonging to $P_G$ and the processes *not* using $n_d$ to reach one of the processes represented by $P_G$. Because the processes belonging to $P_G$ will migrate to the same node, the communication between these processes will take place locally. Therefore, the internal load of $P_G$ cannot affect the migration benefit. For that reason, we factor out from $NL_{sd}^G$ the load between any pair of processes that both belong to $P_G$ (see the negative part of Eq. 8).

*Migration benefit $B_{sd}^G$*: Represents the benefit of migrating a super-process $P_G$ from $n_s$ towards $n_d$, which is given by Eq. 9.

$$PL_{sd}^G = \sum_{\forall i: p_i \in P_G} pl_{sd}^i \qquad \text{Eq. 7}$$

$$NL_{sd}^G = \sum_{\forall i: p_i \in P_G} nl_{sd}^i - \sum_{\forall i: p_i \in P_G} \sum_{\forall k: p_k \in P_G} c_{ik} \qquad \text{Eq. 8}$$

$$B_{sd}^G = PL_{sd}^G - NL_{sd}^G \qquad \text{Eq. 9}$$

*Center of gravity*: Following the respective definition for a single process, we can predicate whether a super-process is at the center of gravity of its communication load or otherwise. However, it is not simple to identify the set of processes that form a super-process that is not at the center of gravity of its communication load. To achieve the above identification, we follow the steps described below.

*Identifying a group of processes forming an unbalanced super-process*: Assuming a pair of nodes ($n_s$, $n_d$: $h_{sd} = 1$). We follow the below mentioned steps to find which super-process must migrate from $n_s$ towards $n_d$ to further reduce the total network cost.

**Step 1:** Construct a graph (called *min-cut graph*) where the nodes represent the *real* processes hosted by $n_s$, while the edges represent the communication dependencies between the processes in question.

**Step 2:** Add $n_d$ to the graph and for each process $p_i$ depicted in the graph draw an edge between $n_d$ and $p_i$, with the weight of such an edge being equal to the positive load $pl_{sd}^i$. The aforementioned weight represents the partial benefit of migrating $p_i$ towards $n_d$.

**Step 3:** Add $n_s$ to the graph and for each process $p_i$ pictured in the graph draw an edge between $n_s$ and $p_i$, with the weight of that edge being equal to the negative load $nl_{sd}^i$ minus the load between $p_i$ and its *real* co-located processes. The above mentioned weight represents the partial cost incurred in case $p_i$ migrates towards $n_d$. The reason we do not take into account the load between $p_i$ and its real co-located processes is that we are not yet ready to identify whether it is beneficial to migrate $p_i$'s co-located processes when considering the migration of $p_i$. However, the above identification takes place when finding the min-cut in the resulting graph (step 5).

**Step 4:** Remove the edges having weight equal to zero. In case the procedure results in a partitioned graph, then the processes belonging to the same partition as $n_d$ are migrated towards $n_d$, and we do not proceed to the next step.

**Step 5:** A min-cut algorithm is applied to the resulting graph, with $n_s$ and $n_d$ playing the role of source and destination, respectively. By doing so, we identify the group of processes (called super-process) that must migrate from $n_s$ towards $n_d$ to maximize the migration benefit. Specifically, the most beneficial super-process is formed by the processes belonging to the same partition as $n_d$. Note that we always perform the migration of $P_G$ towards $n_d$, except for the case where the migration benefit is equal to zero. We also must note that the benefit cannot be less than zero as per the definition of min-cut.

*Min-cut graph example*: Consider the network depicted in Figure 5, which consists of 4 nodes that host an application of 6 real processes. Specifically, $n_2$ hosts {$p_1$, $p_2$, $p_3$, $p_4$}, while $n_3$ and $n_4$ host the processes $p_4$ and $p_6$, respectively. To identify the super-process that maximizes the migration benefit from $n_2$ to $n_3$, we apply the following steps:

**S1**: Construct a graph with the real processes {$p_1$, $p_2$, $p_3$, $p_4$} hosted by $n_2$ and their in-between edges (left side of Figure 6).

**S2**: Draw an edge between $n_3$ (destination node) and each process $p_i$ depicted in the min-cut graph. At the right side of Figure 6, we can see the weight of the edge between $p_2$ and $n_3$, which is equal to $pl_{23}^2 = 7 + 10 = 17$. Note that the processes $\{p_4, p_6\}$ use $n_3$ to reach $p_2$.

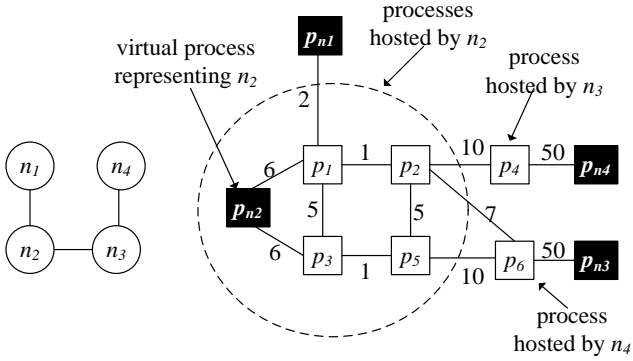

Figure 5. Application example

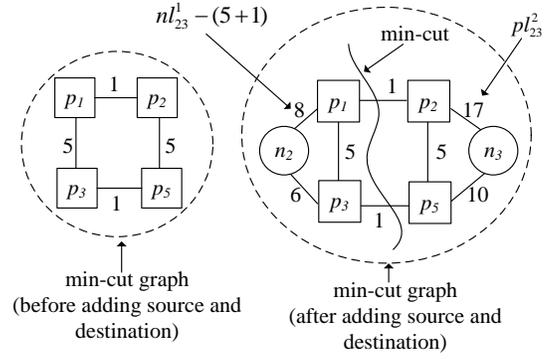

Figure 6. Min-cut graph

**S3**: Add $n_2$ (local node) to the graph and draw an edge between $n_2$ and each process $p_i$, depicted in the min-cut graph. On the right side of Figure 6, we can see the weight of the edge between $p_1$ and $n_2$ that is equal to $nl_{23}^1 - (5+1) = (6+2+5+1) - (5+1) = 8$. Note that "5+1" represents the load between $p_1$ and its *real* co-located processes $\{p_2, p_3\}$.

**S4**: Remove any zero-weighted edge that appears in the min-cut graph. The resulting graph is shown on the right side of Figure 6. As we can see, there is no partition in the graph. Therefore we proceed to Step 5.

**S5**: Applying a min-cut algorithm on the resulting graph reveals that the super-process $P_G = \{p_2, p_5\}$ maximizes the migration benefit from $n_2$ to $n_3$, with $B_{23}^G = 17 + 10 - 2 = 25$. Therefore, by migrating $P_G$ from $n_2$ to $n_3$ we reduce the total network cost by 25.

*C. Super-process Migration Mechanism*

Each node in the system is responsible for migrating or otherwise its hosted processes by invoking the single or super-process migration mechanisms. The single migration mechanism is invoked only in the case where the calling node hosts processes that are not adjacent to each other. Otherwise, the super-process migration mechanism is called to avoid sub-optimal cases, as reported in Figure 3.

**Beneficial migrations.** For a node to predicate whether a single process or super-process migration is beneficial or not, it needs only the knowledge of: **(a)** the volume of data exchanged for the external communication of the processes it hosts, **(b)** the volume of data exchanged for the internal (local) communication of the processes it hosts, and **(c)** the 1-hop neighbors involved for the aforementioned external communication. Because all of the above mentioned information is already known to the nodes within the system, the execution of DRA is based only on local information.

**Dynamic changes.** To capture the dynamic changes within the traffic volume between two processes, we adopt the averaging technique used in [14]. Specifically, we maintain the volume of data exchanged between two processes (called $p_i$ and $p_j$) as an exponential average over its past values:

$$volume_{ij}[t] = a * volume_{ij}[t-1] + (1-a) * c_{ij}[t],$$

where $a$ is the averaging constant with a value between 0 and 1, $volume_{ij}[t]$ and $volume_{ij}[t-1]$ represent the volume computed for the monitoring windows at time $t$ and $t - 1$ respectively, while $c_{ij}[t]$ is the traffic measured at time $t$.

**Oscillations.** Useless migrations may happen when having frequent transitions from one communication pattern to another one. To avoid such oscillations, we adopt the technique of *inertia factor* $\gamma > 1$ used in [14]. Each time a node considers whether a migration is beneficial or not, it multiplies the negative load (in Eq. 6 and Eq. 9) with the inertia factor $\gamma$.

**Hierarchical networks.** DRA has been designed to run in tree-structured networks. However, DRA can be easily extended to run in hierarchical networks. Assume a system of $N$ clusters networked as a tree, with each cluster consisting of $M$ fully connected servers. A server belonging to a cluster (hereafter referred as $C_s$) is allowed to perform an inter-cluster process (or super-process) migration towards an 1-hop neighboring cluster (hereafter, called $C_d$). Such a migration reduces the network cost iff Eq. 6 (or Eq. 9) gives a benefit greater than zero. Note that $s$ and $d$ in Eq. 6 and Eq. 9 play now the role of $C_s$ and $C_d$, respectively. If a server cannot perform any inter-cluster migration, then it considers performing beneficial intra-cluster migrations. However, due to the fact that the servers within a cluster are assumed fully connected, we must change the way we calculate the positive and negative load. Specifically, Eq. 4 becomes

$$pl_{sd}^i = \sum_{k=1}^{P+N}(c_{ik}+c_{ki})f_{dk},$$

which captures the fact that when migrating $p_i$ from server $s$ to server $d$, the positive load is equal to the sum of the communication load between $p_i$ and any process hosted by the destination server $d$. While Eq. 5 becomes

$$nl_{sd}^i = \sum_{k=1}^{P+N}(c_{ik}+c_{ki})f_{sk},$$

which states that when migrating $p_i$ from server $s$ to server $d$ the negative load is equal to the sum of the communication load between $p_i$ and any process hosted by the source server $s$.

**Capacity constraints.** We extend the functionality of DRA to consider capacity constraints on servers. Specifically, if a node has no sufficient storage or CPU capacity to host a super-process, then we resort to prune one or more processes (the less beneficial ones) belonging to that super-process.

## V. CONVERGENCE AND OPTIMALITY

### A. Convergence

In order for DRA to converge in an assignment scheme, it is required that either **(a)** the communication patterns between processes are static; or **(b)** there is a sufficient amount of time between the transitions from one communication pattern to another one in order for DRA to identify each unbalanced process (or super-process). To proceed with the proof of DRA's convergence, we make use of the following lemmas. These lemmas state that the concurrent migrations do not affect DRA's convergence, provided that nodes are not permitted to swap *adjacent* processes in a concurrent fashion. Without loss of generality, from now on for simplicity by saying "process" we will mean "process" or "super-process".

**Lemma 1.** *In case the processes that are migrated concurrently by DRA are not adjacent to each other, then the total network cost is reduced by an amount equal to the sum of their migration benefits (as calculated by DRA).*

**Proof.** It suffices to show that DRA calculates in a correct way the benefits of such migrations. The only case for DRA to miscalculate a migration benefit of a process $p_i$ is to consider that the amount of data exchanged between $p_i$ and an adjacent process to $p_i$ belongs to the negative load, while this amount actually belongs to the positive load, and the opposite (see Eq. 6). The above can happen when the hosting node of $p_i$ is not aware that an adjacent process to $p_i$ is under migration. However, such a miscalculation may happen only in the case DRA is trying to calculate the migration benefit of $p_i$ and at the same time an adjacent process to $p_i$ decides also to migrate. Because in this lemma we do not consider concurrent migrations of adjacent processes, we end the proof. □

**Lemma 2.** *In case two or more adjacent processes are migrated concurrently by DRA, then the total network cost is reduced by an amount equal to the sum of their migration benefits (as calculated by DRA), provided that the distance between each pair of these adjacent processes is greater than or equal to 2 hops.*

**Proof.** It suffices to show that when two adjacent processes $p_i$ and $p_k$ migrate concurrently ($M_{sd}^i$, $M_{s'd'}^k$), then their benefits cannot be miscalculated, provided that the distance of their hosting nodes is at least *2* hops. In other words, the migration benefit of $p_i$ does not depend on whether $p_k$ migrates or not, and the opposite. Assume that $n_s$ calculates $b_{sd}^i$ without knowing whether $p_k$ will migrate or not. Two cases arise: **(i)** the destination node $n_d$ is used from $p_k$ when the latter needs to communicate with $p_i$ (when $p_i$ is located on $n_s$). Consequently, the load between $p_i$ and $p_k$ will belong to the positive load $pl_{sd}^i$ irrespective of whether $p_k$ migrates or not. **(ii)** the destination node $n_d$ is **not** used from $p_k$ when the latter needs to communicate with $p_i$ (when $p_i$ is located on $n_s$). Consequently, the load between $p_i$ and $p_k$ will belong to the negative load $nl_{sd}^i$ irrespective of whether $p_k$ migrates or not. Due to symmetry, the same holds when the hosting node of $p_k$ tries to calculate migration benefit of $p_k$. Therefore, we conclude that for all the above cases the migration benefit of some process $p_i$ does not depend on concurrent migrations of processes adjacent to $p_i$. □

**Lemma 3.** *In case two or more adjacent processes are migrated concurrently by DRA, and assuming that the distance between each possible pair of these adjacent processes is equal to 1 hop; then, the total network cost is reduced by an amount equal to the sum of their migration benefits (as calculated by DRA,) provided that nodes are not permitted to swap adjacent processes in a concurrent fashion.*

**Proof.** Following the same reasoning as in Lemma 2 we result in the same two cases stated in Lemma 2. In terms of case (ii), we conclude that the migration benefits are not miscalculated irrespective of whether $p_k$ migrates or not. Regarding the case (i), the migration benefits are not miscalculated if process $p_k$ decides to migrate to a node other than the hosting node of $p_i$; while they are miscalculated if $p_k$ migrates to the hosting node of $p_i$. The above miscalculation is attributed to the fact that the hosting node of $p_i$ will calculate the migration benefit of $p_i$, assuming that the load between $p_i$ and $p_k$ belongs to $pl_{sd}^i$. However, because $p_k$ will migrate to the old hosting node of $p_i$, it is obvious that the above assumption does not hold true. Due to symmetry, the same holds for the calculation of the migration benefit of $p_k$. Therefore, the migration benefit is miscalculated only when there are swaps of adjacent processes. By not permitting such swaps we end the proof. □

**Lemma 4.** *In case two or more adjacent processes are migrated concurrently by DRA, then the total network cost is reduced by an amount equal to the sum of their migration benefits (as calculated by DRA), provided that the distance between each pair of these adjacent processes is equal to 0 hops (co-location).*

**Proof.** The node is aware of any hosted process to be migrated, therefore the benefits cannot be miscalculated. □

**Theorem 1.** *Assuming an old assignment scheme and some changes in the communication patterns between processes, then DRA will converge into a new assignment scheme by performing a bounded number of migrations.*

**Proof.** In case DRA performs process migrations in a sequential fashion (no concurrent migrations), then it is obvious that the total network cost will be reduced by $b_{sd}^i$ units each time a migration $M_{sd}^i$ takes place. By making use of Lemmas 1, 2, 3 and 4, we also conclude that the above holds true when concurrent migrations take place, provided that swaps between adjacent processes are not permitted. Because DRA does not permit such swaps and due to the fact that the total network cost is represented by a number that is *always* greater than or equal to zero, we conclude that the total number of migrations are bounded by the total network cost (convergence). □

### B. Optimality

The proof for the optimality of our algorithm is restricted to the case where **(a)** no capacity constraints are considered (because otherwise the problem becomes NP-complete [3]); and **(b)** the structure of the network is a tree. Before proceeding with the theorem proving DRA's optimality, we need some auxiliary proofs (Lemmas 5, 6, 7) which depend on the following definition. An assignment scheme whereby each process or super-process is in its center of gravity is named *totally balanced assignment scheme* (TBAS). The basic idea behind these auxiliary proofs is to show that when a TBAS has been achieved, then any other migration or series of migrations cannot reduce further the total network cost.

**Lemma 5**: *Given a TBAS, then any attempt of migrating a process to any node in the system leads always into a new network cost which is greater than or equal to the old one.*

**Proof.** As per the TBAS definition, there is no 1-hop migration of any process in the system leading in total network cost reduction. Therefore, what it remains is to show that the above applies for any *k*-hop migration.

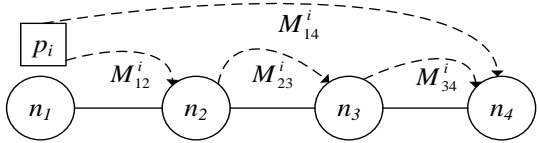

Figure 7.  Decomposing a 3-hop migration into three 1-hop migrations

Consider the example shown in Figure 7. As we can see there is a 3-hop migration $M^i_{14}$ which can be decomposed into three 1-hop migrations ($M^i_{12}$, $M^i_{23}$, $M^i_{34}$). The benefit of $M^i_{14}$ is equal to the sum of the benefits of these three 1-hop migrations: $b^k_{14} = b^k_{12} + b^k_{23} + b^k_{34}$ (Eq. 10). Recall now that the benefit of a process migration can be broken down to the positive load ($pl^i_{sd}$) and the negative load ($nl^i_{sd}$). Looking at the positive loads of these migrations we claim that the following holds $pl^i_{12} \geq pl^i_{23} \geq pl^i_{34}$. Recall that $pl^i_{sd}$ represents the communication load between $p_i$ (located on $n_s$) and the processes using $n_d$ as either a routing or hosting node to reach $p_i$. In our example *A* is called the set of processes using $n_2$ to communicate with $p_i$ (when $p_i$ is located on $n_1$); *B* is called the set of processes using $n_3$ to communicate with $p_i$ (when $p_i$ is located on $n_2$); while *C* is named the set of processes using $n_4$ to communicate with $p_i$ (when $p_i$ is located on $n_3$). We observe that, *A* is a superset of *B*, which in its turn is a superset of *C*. Therefore, $A \supseteq B \supseteq C \Rightarrow pl^i_{12} \geq pl^i_{23} \geq pl^i_{34}$ (Eq. 11). Following the same rationale we prove that $nl^k_{12} \leq nl^k_{23} \leq nl^k_{34}$ (Eq. 12). Combining Eq. 6, Eq. 11, and Eq. 12, we result in the following equation $b^k_{12} \geq b^k_{23} \geq b^k_{34}$ (Eq. 13). The combination of Eq. 13 and Eq. 10 gives the following: $b^k_{14} \leq 3b^k_{12}$. Note that $n_2$ is the 1-hop neighbor of $n_1$, which is used by $n_4$ to reach $n_1$. The above can be generalized into the next statement :*assuming a k-hop migration in a tree-structured network from a node $n_s$ to a node $n_d$, and an 1-hop neighbor (called $n_z$) used by $n_s$ to reach $n_d$, then the following equation holds true*: $b^i_{sd} \leq k * b^i_{sz}$ (Eq. 14).

Getting now back to our case, we recall that there is no 1-hop process migration with benefit greater than zero (TBAS assumption). Note also that $b^i_{sz}$ in Eq. 14 represents the benefit when migrating $p_i$ by 1-hop. Therefore, according to Eq. 14, in the best case any *k*-hop migration will have benefit *k* times the zero. The above brings us to the conclusion that there is no *k*-hop migration with benefit greater than zero. Consequently, there is no *k*-hop migration that can reduce further the total network cost. □

**Lemma 6:** *Given a TBAS, then any migration of a process leaving the total network cost intact or increasing it cannot cause a series of other migrations that eventually reduce further the total network cost.*

**Proof.** To result in a contradiction, we assume that $p_i$ is a process migrating from $n_s$ to an 1-hop neighbor $n_d$, leaving the total network cost intact (or increasing it) but causing other migrations that eventually reduce it. This implies that an adjacent process of $p_i$ (let $p_k$) is affected by $M^i_{sd}$ and becomes unbalanced. We consider now the following three scenarios: **(i)** $p_k$ is located on a node using $n_d$ to reach $p_i$. However, the above entails that $p_k$ was not in its center of gravity before $M^i_{sd}$ takes place (contradiction due to TBAS). **(ii)** $p_k$ is located on a node (other than $n_s$) that does not use $n_d$ to reach $p_i$. However, the above implies that $p_k$ was not in its center of gravity before $M^i_{sd}$ takes place (contradiction due to TBAS). **(iii)** $p_k$ is co-located with $p_i$, which entails that both $p_i$ and $p_k$ form an unbalanced super-process (contradiction due to TBAS). Therefore, we conclude that there cannot exist such an $M^i_{sd}$. To complete our proof, we need to prove that a contradiction holds also true for the case where $n_d$ is a *k*-hop neighbor of $n_s$. Because $M^i_{sd}$ is a *k*-hop migration leaving intact the total network cost (or increasing it), then it is equivalent to say that there are *k* 1-hop migrations from $n_s$ to $n_d$, where each of them leaves intact the total network cost (or increasing it). Applying the first part of the proof for each of these *k* migrations we end the proof. □

**Theorem 2.** *DRA results always in an optimal assignment scheme.*

**Proof.** When DRA converges to a new assignment scheme, then it stops performing migrations (see Theorem 1). If we combine the above with the fact that DRA stops performing migrations only in the case where each process or super-process in the system is in its center of gravity, then we conclude that DRA results always in a TBAS. Therefore, combining the above with lemmas 5 and 6 we conclude that when a TBAS has been achieved, then there is no migration or series of migrations leading in a more beneficial assignment scheme than the TBAS. The above entails that DRA results always in an optimal assignment scheme. □

VI. CONCLUSIONS

In this work, we studied the problem of minimizing the resources consumed by an application during its execution on tree-structured and hierarchical networks. The problem was solved in an optimal way (tree-structured networks) by proposing a fully distributed algorithm guaranteeing convergence. Our future plans include enhancing the functionality of DRA to take into consideration other network structures besides tree and hierarchical networks.